# Misbehavior in Mobile Application Markets

*Security and Cooperation in Wireless Networks Mini-project*

Steven Meyer
EPFL 2010

Supervisor: Julien Freudiger
Professor: Jean-Pierre Hubaux

## 1  Introduction

Smartphones feature advanced computing ability and connectivity compared to basic cell phones. They are notably able to run applications that users can install from mobile application markets. To do so, most smartphone manufacturers maintain a mobile application market operating much like a Debian repository.

Mobile application markets facilitate the distribution of applications and thus help developers advertise their work and customers find useful applications. In addition, the operators of mobile application markets can control the quality and the content of the applications. These markets are growing rapidly with more than 300'000 application in the App Store of Apple and more than 100'000 in the Android Market of Google. This is not only a great opportunity for phone manufacturers to earn money but also for indie developers (single or small teams of developers with small financial support) who can thus have a great distribution channel. Steve Demeter, the Trim game developer for iPhone, became millionaire with a single puzzle game[1].

Obviously, as new markets generate a lot of money, the temptation of misbehavior to steal part of the benefits is big. The first famous case was the one of *Molinker*[2] who self-rated his applications with 5 stars to pump up his ranking in order to increase its revenue stream. Also, in summer 2010, Thuat Nguyen[3] used stolen credit cards and iTunes accounts to buy his own books: this earned him a lot of money but, as he ranked 42 of his books in the top 50, the operation became suspicious to Apple. Finally, the Aura Faint application was removed from the App Store because it was uploading all the contacts of the phone to the developers' server: it is unknown how exactly this information was to be exploited, but it certainly could have been very useful for spamming.

In this report, we will consider the problem of misbehavior in mobile application markets. We will investigate multiple attacks by misbehaving developers, users or network operators that aim at breaking rules for their own benefit, managing to outwit the operators' control on which applications can be installed. We notably suggest novel attacks that may affect mobile markets in the future: in particular, we show that it is possible to get revenue for applications created by someone else, trick a user to download and buy an application and new ways to pump up an application's ranking. We will also discuss possible solutions against spyware applications and cheating developers.

---

[1] http://www.wired.com/gadgetlab/2008/09/indie-developer/

[2] http://www.engadget.com/2009/12/08/molinker-is-no-more-on-the-app-store-ratings-scam-results-in/

[3] http://voices.washingtonpost.com/fasterforward/2010/07/apple_boots_iphone_developer_o.html

## 2 Related Work

Seriot (1) showed that even though iPhone applications were sandboxed and that only a few APIs were available in the SDK, a malicious developer could access more information than expected. He created a simple proof of concept application that could extract all contacts and many past GPS positions without the user seeing a warning. He concluded that even though the App Store validation process is very strict, malware is still able to pass through.

Enck et al. (2) worked on the TaintDroid application, which tracks the data flow in an Android application, to analyze how private data is manipulated by third-party applications. After analyzing 30 of the top popular applications, their results showed that, two-thirds of them exhibited suspicious handling of sensitive data that might have led to privacy issues.

Smith (3) shows that the unique ID of iPhones is available to developers and highlights problematic privacy consequences. For example, online services could track users using such identifier; he concludes that this is an important loss of anonymity as services could link pseudonyms with real users.

Girardello and Michahelles suggested an alternative application market (4), where the ranking would be based on usage and location rather than on users' vote. The advantage of their solution compared to the existing rankings is that it would take into account more real-life parameters to enhance the user's experience.

## 3 Mobile applications security model

We start by describing the different solutions put in place on smartphones to limit misbehavior by third-parties without affecting users' experience.

### 3.1 Mobile application markets strategies

In this chapter, we compare two mobile application markets and see the different models that they put in place to distribute applications.

The App Store by Apple[4] is the most popular with more than 300'000 applications available. The only way for a user to install applications on his iPhone (unless the iPhone is jailbroken) is to download them from the App Store. In order to publish an application in the App Store, Apple must first certify developers: their identity is verified and an annual fee must be paid. Then, their applications must be verified and approved by Apple; they check for example the content of the application and the data accessed. This way, Apple controls which applications are installed on the phone and thus provides a layer of security for end users.

---

[4] http://www.apple.com/iphone/features/app-store.html

Sandboxing is used and will be described later.

The Android Market[5] is the default market on the Android phones; it is owned by Google and has more than 100'000 applications. The developers are not certified, nor are the applications verified before being published, but rather removed due to complaints. The restrictions to publish an application in the Android market are roughly the same as on the App Store (except that there is no subjective refusal of an application). But since Android is an open platform, Android Market does not have the exclusivity. Anyone can create its own market and deploy it on Android phones (although to install a new market one should be an experimented user, as it is not possible to find third party markets on the default Android Market. One would thus need to find third party markets and download them through the browser). Therefore, there is absolutely no control on applications deployed on an Android device and malware might easily be installed.

## 3.2 Markets on Android

Since the Android phone enables users to install third party application markets, it is interesting to look at the differences and advantages between them.

We can see four main differences in the markets on Android (it is important not to confuse "the Android Market", which is the official market on Android, with the "market on Android", which could be any market installed on the platform).

**The fee to submit an application**: to be efficient, a market needs human resources that will not only maintain and supervise it but that will also verify the content and quality of the applications that are submitted. (On some platforms, part of the fee is also used to generate a certificate for the developer that will be used to sign the application with the help of a third party such as VeriSign: but on Android, the applications can be self-signed, therefore there is no need to be certified by a central authority).

**The income cut**: owning an application market on a mobile platform is like owning a goose that laid the golden eggs. Most of the mobile application markets take a percentage on every transaction, therefore the more successful is a market and the more paid apps are downloaded, the higher the income is. The income cut is generally the business opportunity in the application market model.

**Application approval**: the process of accepting an application can vary a lot between different markets, going from human and automated testing to 100% acceptance without verification. Generally, the approval process defines the quality of a market and the risk of finding malware.

**Rating, ranking and promotion**: this is one of the most important criteria for a developer. The rating is generally done by the end users, the ranking is done by the market operator with the input of the rating and finally the promotion is done by the market operator.

The main portal is Android Market: it is the default one owned by Google and preinstalled in all Android devices. It now offers more than 100'000 applications. To be able to submit an application on this portal, the developer must first pay a 25$ fee. Google's additional income is 30% of the sale price. Even though Android market accepts a priori all applications, some might be processed to ensure that they follow the guidelines and policies (we can find restrictions such as nudity, unpredictable network usage, harm to the devices or its data).

---

[5] http://www.android.com/market/

| Name | Website | # apps | Enter fee | Operator's cut per sale | Restriction on app | Ranking | Sail argument |
|---|---|---|---|---|---|---|---|
| Android Market | android.com/market | 100'000 | 25$ | 30 | No: nudity, unpredictable network usage, harm devices or data | Users to rate Products + Android Stuff rating | Main market of Android |
| Appslib | appslib.com | NA | 0$ | 30 | Run on device & interest to an end-user. | Operators + end users | App for non standard Android devices |
| MiKandi | mikandi.com | NA | 0$ | 35 | Remove on complain | users to rate Products + MiKandi Stuff rating | Open to all kind of software |
| SlideME | slideme.org | NA | 0$ | 5 | NA | NA | Better promotion (website store: video chat etc.) |
| androidgear | androidgear.com | NA | NA | 45 | NA | NA operator marketing only | newsletter & 1st level end-user support |

**Figure 1: application markets on Android**

MiKandi[6] is another interesting market, which positions itself as the most open one. There is no entry fee, no restriction on the quality nor the content of an application (therefore we find a lot of nudity related applications available) and they are not verified (nevertheless an application can be removed due to complaints). The catches with this market are the sales cut of 35% for the market operator and the general distrusted reputation of the applications.

There are many other portals such as AppsLib[7], SlideMe[8] and AndroidGear[9], whose main differences are controls carried out on the applications, the portal fee cut and the promotion provided to the applications. See Figure 1: application markets on Android

### 3.2.1 Sandboxing

The sandboxing is a general IT principle that became a standard to cope with third-party applications in the smartphone environment. A sandboxed application is totally isolated from the rest of the system and the other applications and it can only communicate through predefined APIs that are controlled by the OS. The purpose of this system is to ensure that an application cannot harm the system and to avoid private information abuse from the phone or from other applications. On Android, applications are totally isolated from each other (except if they are written by the same developer who could create connections between his apps), but there are no limitations on the APIs: an application with the correct permission could therefore even *brick* the phone (transform the phone into an unusable brick). All the resources needed by the application are written in a manifest file that that will prompt the user with a warning message during the installation, giving to the user the choice of either continuing the installation or aborting it(5).

---

[6] http://www.mikandi.com/
[7] http://appslib.com/
[8] https://slideme.org/
[9] http://www.androidgear.com/

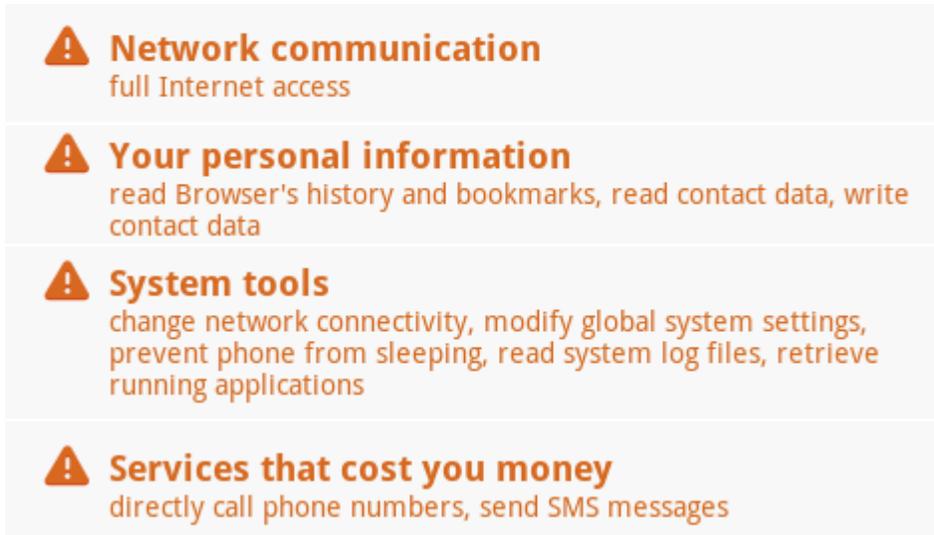

Figure 2: privilege warning messages

Android has 115 different permissions[10] that are categorized into a dozen groups to make them more understandable for the user: for example a category "that costs money" contains sending SMS and outgoing phone calls. However, this specific structure might be a problem in itself as it puts many different permissions under the same banner. This could mislead the users into installing modifications of the system along with the application: for example in the "system tools" category we can find "prevent the phone from sleeping", which could be legitimate for a game, but also "modify global system settings", which should only be used by very specific applications. Therefore, it relies on human ability to understand what data is shared and this is a potential weakness. See Figure 2

### 3.2.2 Verification

The Android market doesn't verify the submitted applications and relies on the users to report inappropriate content or problematic applications. Nevertheless, Android Market reserves the right to verify an application before accepting it.

### 3.2.3 The Ranking

The ranking process is obscure and not documented in order to avoid manipulation. However, the following assumptions can be done on the different criteria used for the ranking:

- **Number of downloads** is obviously a trivial criterion.

---

[10] http://developer.android.com/reference/android/Manifest.permission.html

- **"Install base"** corresponds to the number of users who have installed the application, then excludes the users that downloaded the application several times or that have uninstalled it.
- **Acceleration:** the ranking would also try to promote new applications with a lot of potential and with a sudden increase in downloads.
- **Velocity:** number of constant downloads could help to detect applications that will become trendy and that should be well ranked.
- **Users' feedback** might affect the ranking or at least influence other users' downloads, which will have some bearing on the other criteria above.

Baptiste Benezet form faberNovel explains the way the App Store from Apple ranking algorithm works: "The formula is 8 times the sales of the current day + 5 times the sales on the 2 proceeding days + 2 times the sales on initial date"[11]. So $Points = 8 \cdot j + 5 \cdot j_{-1} + 5 \cdot j_{-2} + 2 \cdot j_{-3}$

Nevertheless, we can find on the Android forum some complaints about a surprising ranking for a Spanish application situated in the top paid applications, that has a rating of 4.2 and less than 1000 download for the last 2 months[12]. It would be reassuring to know that Google does not distort the ranking of the applications and that official or independent organizations could investigate its search ranking algorithm.

### 3.3 Adversary Model

Before examining the possible types of developer misbehavior, let's first consider who this developer could be and his motivation. In most cases, he will be a malicious developer driven by a financial incentive. His aim will be to earn more money than he is entitled to. He can lure the user into spending more than expected by asking him to pay for a service when the application is free; he can add hidden costs to the application by sending out text messages or phone calls to highly charged services; he can damage the phone and offer an expensive solution to repair it; he can use or steal information from the phone in order to sell it to statistical analyzers (used for publicity for example) and to spammers; he can blackmail the user in exchange of not publishing private information on the internet; finally he can lie to the user by changing his review and ranking for increasing the number of downloads (and payments) of an application that is not worth it(6) (7).

Additionally, there are non-malicious developers who simply desire to harm the user by damaging the phone. Their motivation is not financial, but a pursuit of power, fun, challenges and testing the system. They usually are teenagers who do not fully realize the consequences of their acts.

Finally, we should not forget the developer that has good will but unfortunately his application is damaging the user's phone with bugs, by creating security holes or enabling privacy leakage, such as opening network

---

[11] http://www.readwriteweb.com/start/2010/02/iphone-appstore-ranking-algorithm.php
[12] http://www.google.com/support/forum/p/Android+Market/thread?tid=5707daceb71954de&hl=en

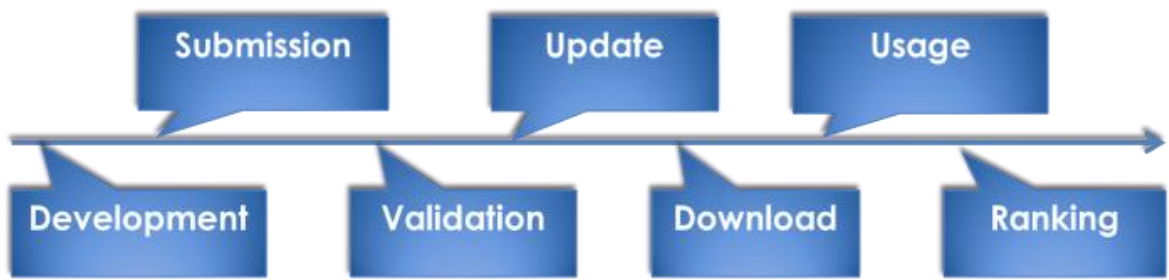
Figure 3 : Life of applications

ports on the phone or uploading contact lists to an unsecured server.

## 4 Application life cycle

We will now have to look at the different steps in the lifecycle of an application in order to understand how to trick the system or successfully install malware on an Android phone. See Figure 3

1. **Development**: the developer implements the functionalities and decides which resources his application will need (such as address book, GPS, camera, core system access).
2. **Submission**: the application is uploaded with a description, keywords and screenshots to the application market. This information will be used by the end users to help them decide if they want to download and use the application.
3. **Validation**: it does not exist in every market, since some markets accept applications without verifying them before, but, when done, it can last 2 to 7 days.
4. **Update**: where the developer can change the functionalities, description and price of his application.
5. **Download**: hopefully the application is downloaded and the developer can see its statistics and get revenue.
6. **Usage**: the user will launch the application and use it or let it work in the background.
7. **Ranking**: finally, after the application is used, the user gives a raking to the application and adds comments to help other users decide if the application is worth it.

## 5 Misbehavior

In this section, we consider potential attacks on each phase of mobile application lifetime.

### 5.1 Development phase

As suggested in paper (1), during the development phase, the developer would try to trick the system by fooling the sandbox or by abusing some of the APIs. An application in Android is not totally sandboxed; it can communicate with other processes or with the hardware. To be able to exit from the sandbox, the developer specifies in a manifest which resources will be needed by the application; then, during the installation, the user will have a warning message asking approval for the application to use the resources. If a developer is able to access re-sources without going through the expected API and therefore not specifying them in the manifest, then he can avoid having the warning message displayed on the phone: this will fool the system by accessing resources on the phone it is not supposed to.

## 5.2 Submission phase

In the section, we suggest a new type of attack.

One peculiarity of the android platform is that we can find many different markets. This is a great advantage for developers who don't wish to follow strict rules or who want to benefit from different promotion and distribution methods. However, since there are multiple markets and the developer self-signs his applications, the strength of openness could turn into a weakness: someone could download an application (from another developer) and impersonate it as if he was the author. The steps to do so are the following:

1. Be rooted on the device by either installing a custom Rom form, for example, XDA-developer[13] or by *Rooting* the OS[14].
2. Install a file explorer that allows root access such as *Astro File Manager*[15] or equivalent to get access to package of the download applications from Android market.
3. Find, download and install the application chosen to be impersonated from one market.
4. Extract the application from "data\apps" or "data\pvtapps" repertory on the phone onto the computer.
5. Create a developer account in the other application market for Android.
6. Publish the application as if it was his.

If it is published for a lower price than the one published by the real developer and on many markets, it might even be downloaded more often than the genuine one and therefore generate even more money.

## 5.3 Validation phase

The Android Market does not verify nor validate submitted applications. Instead, it relies on user flagging and complaints to check if an application might harm the phone. Then, depending on complaints, it verifies the application and possibly decides to remove it. A way to remove a concurrent application could be to flag it as inappropriate, then Google might remove it from the market during the tests (but the removal would be for less than a day so not very effective).

## 5.4 Update phase

In the section, we suggest a new type of attack.

In this phase we are going to discuss social engineering. The aim is to trick existing users of a good application to update to a new version that contains malware. The developer creates a nice application that will be downloaded for its functionalities (whatever they are) then there will be an update which will contain the malicious code. There will obviously be a warning that will appear to the user during the installation phase; but, as the user liked the application (and trusts the developer because his first version worked well), he will simply discard the warning and install the update of the application.

---

[13] http://forum.xda-developers.com/
[14] http://www.ryebrye.com/blog/2009/08/16/android-rooting-in-1-click-in-progress/
[15] http://www.metago.net/astro/fm/

Another possible misbehavior is to trick the users into downloading expensive applications by making the users believe they are worth it. In order to achieve this, the developer should first submit a nice and simple application for free. After a while, when it will have been downloaded many times, the ranking will be good and the reviews positive. At this point the developer will change the description of the application by adding many great new features (that don't exist), change the screen shots with the fake functionalities added, and obviously charge an expensive price. Existing users will then not only want to update and pay, but also new users will rely on the great results of the first version to decide to buy the application.

## 5.5 Download phase

Several attacks described in this paper also happen during the download phase, such as the Sybil or the impersonation attacks.

## 5.6 Using phase

This is the phase where the malware will actually be executed. Since on Android there are no private APIs (like on iPhone), if the user blindly accept all the permissions of an application, the application will have full control of the phone and be able to harm it. At this point, we could find exactly the same malware on an Android phone as on a PC, such as key loggers, phishing applications, spamming etc. (7) (6).

## 5.7 Ranking phase

The ranking position of an application is one of the most important criteria of its success. An obvious way to misbehave and pump up an application ranking is to use a Sybil attack (8) and create many Gmail accounts that will be used to fake several users who will then all download and rate the same application.

To do so, we have used an HTC HD2 running on Windows mobile with an Android virtual machine. This way we could totally wipe clean the device between "users" and no serial number of the device could be sent to Android Market. The process of cleaning up the device and rebooting the virtual machine took roughly 15 minutes. To create a Gmail account, one needs to provide a telephone number that will be used to communicate a onetime passcode. This complicates the attack since it might require many different phone numbers to avoid being detected by Google. We therefore created instead, during the setup of the phone, accounts that only require a name, email, password, a secret question and a Capcha. Filling up the form and accepting the terms took approximately 2 minutes.

Then we had to find an application with very few downloads and no rating to be able to influence it as much as possible (it is not easy to find this kind of application since they usually don't appear in the top 40 even when using the search by name option). We decided to use the search keyword "Marge": out of the 37 results we chose the application in the 15[th] position "*Large marge from peewees big A*" by "Metro Videomayors" (this application really looks like malware and we don't advise you to use it) that had no ranking, no comments, and less than 24 downloads. At every iteration of the Sybil attack, we downloaded the application, put 5 stars rating, added a positive comment and tagged all the other comments as useful. This experience was carried out 26 times and the ranking of the application went from 15/37 to 2/37.

These are the steps to follow in order to affect artificially the ranking:

1. Install an Android boot loader on a Windows Mobile device.
2. Boot on Android.
3. Select a language (the same every time).
4. Open the Market application.
5. Create a new Gmail account with a fake name.
6. Find the application to pump up
7. Mark all the existing comments as "Useful"
8. Download the application.
9. Put a 5 star rating with a positive comment.
10. Restart the phone on Windows Mobile.
11. Reinstall the Android boot loader.

This algorithm could be accelerated by an automated script on an Android simulator, where the only user interaction would be to enter the Captcha at the creation of an account, and therefore it could generate an arbitrary large number of votes.

# 6 Possible Countermeasures

In this section we will suggest some measures that the Android team, the Android Market team and the users could implement to avoid the above attacks.

## 6.1 Development phase

This counter measures are aimed at the Android development team. It is obvious that the team is under high pressure to come out every few months with a new version of the OS and with new functionalities (9). It is important that at every release they ensure that the access to each resource is exactly limited to one and only one privilege; this way they could for example avoid giving GPS coordinates by reading the metadata from the pictures.

## 6.2 Submission phase

This point is very tricky because it would necessitate collaboration between different market providers that are competitors. This collaboration would require some efforts and might reduce their income, since they would have to refuse illegitimate application (though it would protect the developers).

There are two possible solutions to this problem. The first one would be to ask the developer to prove the ownership of the source code. This could be done, for example, by requiring the developer to submit the application twice: once with the application that is supposed to be submitted to the market, and the second time adding to the application a given nonce that would appear on the screen. The disadvantages of this solution are that it would be time-consuming for the developer and would require the market operator to manually verify every application submitted (which is not done at the time being).

The second possibility would be for the market to verify whether a new application has already been submitted on another market and, should this be the case, if it is has been submitted by the same developer. Let's suppose that the Android platform could offer a way for all the markets to communicate between them. First one should be able to verify if an application already exists on the market. Using a hash cannot suffice since any insignificant modification to the application (such as changing the version number or the developer's name in the metadata with a hex tool) would totally change the resulting hash. Verifying the signature would also not be enough, as the malicious developer could simply replace the real developer's signature with his own.

This problem of impersonation is atypical, as usually the malicious person tries to pretend that the data he provides originates from somebody else, and not that someone else's data is coming from him.

One advice that we could give to the developer is to clearly display his identity in the application: however, even this doesn't insure that the developer will receive the payment for the purchased application, as the billing is done by the market itself and is not controlled by the application, but at least in case of litigation he would easily be able to prove ownership of the application.

### 6.3 Validation phase

The Android Market team should definitively validate applications put on their market, even though on the Apple's App Store, which is doing validation, we can find malware.

If one market on Android would wish to stand out, its best option would be to instate some kind of validation and then guarantee less malware on its market.

### 6.4 Update phase

An obvious way to protect the users from the misbehavior in the update phase would be to instate validation, which the application would have to go through at every submission or update.

Another option is to consider every update as a new submission and assign the rating of an application to its version and not to the application itself, but this solution would discourage developers to update their application.

Finally, since the attack relies on social engineering, the defense could work on the same level, warning the user about the differences between the two versions, such as differences in the permission, the description and in the rating.

### 6.5 Using phase

Since the malware acts on the phone the same way it acts on a computer, it would be logical to use the same protective tools on both devices. iPhone somewhat works this way since it requires privileged during usage and not during installation. Even though repeated warnings annoy users, it should be essential for them to be able to choose for every application from the following *firewall* policies: "warn me for every access", "warn me only once" or "I trust this application, don't warn me". Additionally, data logging should be available for the user to be able to see the activity done by the application, such as the number of SMSs sent, the web-sites contacted or the data read from the shared memory (such as photo, music etc.).

Since there are so many application that are so easily downloaded, Antivirus that don't only rely on signature (as they currently do) (9) should become a standard, and analytic sandboxes should be available for testing and proofing applications (10).

### 6.6 Ranking phase

As always with the Sybil attack the problem is the *weak pseudonyms*. A simple way to protect the system against this kind of misbehavior would be to make the creation of a new account more complicated or to use *strong pseudonyms* (8) such as a telephone number or the phone's serial number to verify the user's identity.

The Android Market should use (and might already do) some sort of pattern recognition that could raise a red flag when an account is created, only downloads a specific application, gives it a great rating and never reconnects again.

## 7  Conclusions

In this paper we have seen for every development phase the misbehavior that a developer could do. Some of the attacks, such as the sandbox escape, can already be found in literature, but other attacks, such as the developer impersonation, the multiple download and ranking and the social engineering in updates, have been presented here for the first time.

The same way firewalls became standard in computers, *data filters* and Sandboxes will have to become standards in the smartphone environment. People want to trust the application offered by application markets and want to be able to download applications even from indie developers, without having to fear for malware.

The certification phase should become a standard for market acceptance in order to minimalize the risk of finding malware in applications.

For a platform to be successful in the long run it is important on the one hand for developers to trust the market and know that their revenue won't be stolen, and on the other hand for the end user to know that it can trust the application he installs, that it won't steal his private information nor have hidden costs.

In the future, we intend to verify specifically for every market on the Android which one of the above mentioned misbehaviors could occur and develop; how the markets will react to protect the developers and consumers; whether developers will themselves require added features from the markets in order to be protected and protect the users, and whether the users' demands of safe applications will be answered by new standards in smartphones; and which particular market will eventually be the most secure and reliable for the user and developer.